\documentclass[iop,revtex4]{emulateapj}
\usepackage{graphicx,color}
\usepackage{amssymb}
\usepackage{amsmath}
\usepackage{threeparttable}
\usepackage{lscape}
\usepackage{longtable}
\usepackage{hyperref}
\hypersetup{colorlinks   = true, citecolor    = blue}

\mathchardef\mhyphen="2D

\newcommand{\ovi}{O\,{\sc vi}}

\newcommand{\feii}{Fe\,{\sc ii}}

\newcommand{\siiv}{Si\,{\sc iv}}

\newcommand{\cii}{[C\,{\sc ii}]}

\newcommand{\civ}{C\,{\sc iv}}
\newcommand{\mgii}{Mg\,{\sc ii}}

\newcommand{\angstrom}{\text{ \normalfont\AA}}

\mathchardef\mhyphen="2D

\definecolor{red}{rgb}{0.75,0.0,0.0}
\definecolor{blk}{rgb}{0.0,0.0,0.0}
\definecolor{yel}{rgb}{0.65,0.65,0.0}
\definecolor{grn}{rgb}{0.0,0.75,0.0}
\definecolor{blu}{rgb}{0.0,0.0,0.75}
\definecolor{gry}{rgb}{0.75,0.75,0.75}
\def\Nion{\ifmmode N_\mathrm{\scriptstyle ion} \else $N_\mathrm{\scriptstyle ion}$\fi}

\def\ly{$\lambda$}
\def\hi{H\,{\sc i}}

\def\cii{C\,{\sc ii}}

\def\civ{C\,{\sc iv}}

\def\niv{N\,{\sc iv}}

\def\oiv{O\,{\sc iv}}
\def\ov{O\,{\sc v}}
\def\ovi{O\,{\sc vi}}

\def\neviii{Ne\,{\sc viii}}

\def\nex{Ne\,{\sc x}}
\def\naix{Na\,{\sc ix}}

\def\mgx{Mg\,{\sc x}}
\def\mgxii{Mg\,{\sc xii}}
\def\mgii{Mg\,{\sc ii}}
\def\alii{Al\,{\sc ii}}

\def\siiv{Si\,{\sc iv}}

\def\sixiii{Si\,{\sc xiii}}
\def\sixiv{Si\,{\sc xiv}}

\def\arvi{Ar\,{\sc vi}}
\def\arvii{Ar\,{\sc vii}}
\def\arviii{Ar\,{\sc viii}}

\def\feii{Fe\,{\sc ii}}

\def\nh{\ifmmode n_\mathrm{\scriptscriptstyle H} \else $n_\mathrm{\scriptscriptstyle H}$\fi}
\def\ne{\ifmmode n_\mathrm{\scriptstyle e} \else $n_\mathrm{\scriptstyle e}$\fi}
\def\Qh{\ifmmode Q_\mathrm{\scriptstyle H} \else $Q_\mathrm{\scriptstyle H}$\fi}
\def\Uh{\ifmmode U_\mathrm{\scriptstyle H} \else $U_\mathrm{\scriptstyle H}$\fi}
\def\Nh{\ifmmode N_\mathrm{\scriptstyle H} \else $N_\mathrm{\scriptstyle H}$\fi}

\def\Zsun{\ifmmode {\rm Z}_{\odot} \else Z$_{\odot}$\fi}
\def\Msun{\ifmmode {\rm M}_{\odot} \else M$_{\odot}$\fi}
\def\kms{\ifmmode {\rm km~s}^{-1} \else km~s$^{-1}$\fi}
\def\Lya{\ifmmode {\rm Ly}\alpha \else Ly$\alpha$\fi}
\def\Lyb{\ifmmode {\rm Ly}\beta \else Ly$\beta$\fi}
\def\Lyg{\ifmmode {\rm Ly}\gamma \else Ly$\gamma$\fi}
\def\Lyd{\ifmmode {\rm Ly}\delta \else Ly$\delta$\fi}
\def\neaod{\ifmmode n_\mathrm{\scriptscriptstyle AOD} \else $n_\mathrm{\scriptscriptstyle AOD}$\fi}
\def\necrit{\ifmmode n_\mathrm{\scriptstyle cr} \else $n_\mathrm{\scriptstyle cr}$\fi}
\def\ncr{\ifmmode n_\mathrm{\scriptstyle cr} \else $n_\mathrm{\scriptstyle cr}$\fi}
\def\nepi{\ifmmode n_\mathrm{\scriptscriptstyle PI} \else $n_\mathrm{\scriptscriptstyle PI}$\fi}
\def\gtorder{\mathrel{\raise.3ex\hbox{$>$}\mkern-14mu\lower0.6ex\hbox{$\sim$}}}
\def\ltorder{\mathrel{\raise.3ex\hbox{$<$}\mkern-14mu\lower0.6ex\hbox{$\sim$}}}

\newcommand{\SSS} {SSS}

\newcommand{\Comp}{System}

\slugcomment{Submitted to ApJS 2019 Jul 14; Accepted 2019 Oct 6}

\shorttitle{ }
\shortauthors{Xu et al.}
\shortauthors{}

\begin{document}


\title{HST/COS observations of quasar outflows in the 500 -- 1050\angstrom\ rest-frame: IV.\\ The largest Broad Absorption Line Acceleration}


\author{
Xinfeng Xu\altaffilmark{1,$\dagger$},
Nahum Arav\altaffilmark{1},
Timothy Miller\altaffilmark{1},
Gerard A. Kriss\altaffilmark{2},
Rachel Plesha\altaffilmark{2},
}

\affil{$^1$Department of Physics, Virginia Tech, Blacksburg, VA 24061, USA\\
$^2$Space Telescope Science Institute, 3700 San Martin Drive, Baltimore, MD 21218, USA\\
\hspace{07mm}\
}

\altaffiltext{$\dagger$}{Email: xinfeng@vt.edu}

\begin{abstract}
We present an analysis of the broad absorption line (BAL) velocity shift that appeared in one of the outflow systems in quasar SDSS J1042+1646. Observations were taken by the \textit{Hubble Space Telescope/Cosmic Origin Spectrograph} in 2011 and 2017 in the 500 -- 1050\angstrom\ rest frame.  The outflow's velocity centroid shifted by $\sim$ --1550 km s$^{-1}$ from --19,500 km s$^{-1}$ to --21,050 km s$^{-1}$ over a rest-frame time of 3.2 yr. The velocity shift signatures are most apparent in the absorption features from the \neviii\ \ly\ly770.41, 780.32 doublet and are supported by the absorption troughs from \ov\ \ly 629.73 and the \mgx\ \ly\ly609.79, 624.94 doublet. This is the first time where a quasar outflow velocity shift is observed in troughs from more than one ion and in distinct troughs from a doublet transition (\neviii). We attribute the velocity shift to an acceleration of an existing outflow as we are able to exclude photoionization changes and motion of material into and out of the line of sight as alternate explanations. This leads to an average acceleration of 480 km s$^{-1}$ yr$^{-1}$ (1.52 cm s$^{-2}$) in the quasar rest frame. Both the acceleration and the absolute velocity shift are the largest reported for a quasar outflow to date. Based on the absorption troughs of the \ov*\ multiplet, we derive a range for the distance of the outflow ($R$) from the central source, 0.05 pc $<$ $R$ $<$ 54.3 pc. This outflow shows similarities with the fast X-ray outflow detected in quasar PG 1211+143. We use the acceleration and velocity shift to constrain radiatively accelerated active galactic nucleus disk-wind models and use them to make predictions for future observations.

\end{abstract}


\section{Introduction}
\label{sec:Intro}

Broad absorption line (BAL) outflows are seen in quasar spectra as wide, blue-shifted absorption troughs \citep{Weymann91}. These outflows can reach velocities up to $\approx$ 0.2c and have widths up to tens of thousands of km s$^{-1}$. The outflows provide an important means of carrying energy and mass out of the quasar's central region. Therefore, they likely participate in the interactions between the supermassive black holes (SMBHs) and their host galaxies (see elaboration in section 1 of Arav et al. 2019, submitted to ApJS, hereafter Paper I).



BAL troughs have been commonly observed to have variability on multi-year and shorter timescales, e.g., \cite{Filiz13} reported that 50 -- 60\%  of \civ\ and \siiv\ BAL troughs were found to vary in their survey. However, reported cases of accelerating outflows are much rarer. Detections of BAL acceleration in individual objects have been known for two decades \citep[e.g.,][]{Vilkoviski01, Hall07}. These studies reported that the outflow velocities shifted by up to $\sim$ 70 km s$^{-1}$ over rest-frame times of 1--5 yr and had an acceleration range between 0.03 and 0.15 cm s$^{-2}$. A velocity shift is measured directly from the spectra (e.g., from \civ\ absorption trough centroids) of two epochs, and the average acceleration is calculated by dividing this velocity shift by the quasar rest-frame time between the two epochs. 

Surveys of BAL variability find no clear evidence for accelerating outflows \citep[][]{Gibson08, Gibson10, Capellupo12}. The systematic investigation of \civ\ BAL acceleration/deceleration reported in \cite{Grier16} shows a low detection rate of accelerating outflows (2 out of 140 quasars), where their two acceleration candidates show velocity shifts of up to $\sim$ 900 km s$^{-1}$ over rest-frame times of 3--5 yr. Studies of BAL accelerations are challenging due to several reasons: (1) the need for long time baselines to observe the accumulated small acceleration signatures; (2) the difficulties in disentangling the velocity-dependent line profile changes from a true acceleration \citep{Arav99a}; and (3) the self-blending of BAL troughs \citep[e.g.,][]{Arav01a,Scott14}.  

Even though the observations of BAL acceleration are rare, they can provide valuable constraints on dynamical models of quasar outflows \citep[e.g.,][]{Grier16, Misawa19}, including: radiative driving \citep{Murray95}, magnetic driving \citep{Everett05}, and thermal driving \citep{Balsara93}. 

In this paper, we present the discovery of a velocity shift for a BAL outflow seen in quasar SDSS J1042+1646. We attribute the velocity shift to an acceleration of an existing outflow as we are able to exclude photoionization changes and motion of material into and out of the line of sight (LOS) as alternate explanations (see section \ref{sec:BALacc}). The structure of the paper is as follows. In section 2, we discuss the observations and data. We present the evidence of the outflow velocity shift in section 3. In section 4, we show the photoionization analysis of the outflow. In section 5, we discuss possible causes for the observed velocity shift and compare our results to previous studies. We also use the acceleration and velocity shift to constrain radiatively accelerated active galactic nucleus (AGN) disk-wind models in section 5 and use them to make predictions for future observations. In section 6, we summarize our results. We adopt a cosmology of H$_{0}$ = 69.6 km s$^{-1}$ Mpc$^{-1}$, $\Omega_m$ = 0.286, and $\Omega_{\Lambda}$ = 0.714, and we use the Ned Wright's Javascript Cosmology Calculator website \citep{Wright06}.


This paper is part of a series of publications describing the results
of Hubble Space Telescope (HST) program GO-14777, which observed quasar outflows in the EUV500 using the Cosmic Origin Spectrograph (COS).\\
Paper I \citep{ara20a} summarizes the results
for the individual objects and discusses their importance to various
aspects of quasar outflow research. \\
Paper II \citep{xu20a} gives the full
analysis for 4 outflows detected in SDSS J1042+1646, including the
largest kinetic luminosity ($\dot{E}_k$ = $10^{47}$ erg s$^{-1}$) outflow measured to date
at $R=800$~pc, and an outflow at $R=15$~pc. \\
Paper III \citep{mil20a} analyzes 4 outflows
detected in 2MASS J1051+1247, which show remarkable similarities, are
situated at $R\sim200$~pc and have a combined $\dot{E}_k=10^{46}$ erg
s$^{-1}$.  \\
Paper IV is this work.\\
Paper V \citep{mil20b} analyzes 2 outflows
detected in PKS 0352-0711, one outflow at $R=500$~pc and a second
outflow at $R=10$~pc that shows an ionization-potential-dependent
velocity shift for troughs from different ions.\\ 
Paper VI \citep{xu20c} analyzes 2 outflows
detected in SDSS 0755+2306, including one at $R=1600$~pc with
$\dot{E}_k = 10^{46}$ -- 10$^{47}$ erg s$^{-1}$. \\
Paper VII \citep{mil20c} discusses the other
objects observed by program GO-14777, whose outflow characteristics
make the analysis more challenging.\\



\section{Observations and Data Reduction}
\label{sec:Data}

SDSS J1042+1646 (J2000: R.A. = 10:42:44.24, decl. = +16:46:56.14, z = 0.978 ) is 1 of 10 objects targeted by our HST program GO-14777 (PI: Arav; see Paper I). Observations were taken in 2017 November using the COS G130M and G160M gratings \citep{Green12}. Previous observations were done at a roughly five times lower spectral resolution using the COS G140L grating in 2011 June. The wavelength calibrations of these gratings are described in the COS Instrument Science Report (2010-06)\protect{\footnote[1]{\protect\url{http://www.stsci.edu/hst/cos/documents/isrs/ISR2010_06.pdf}}}, where the specified 1$\sigma$ wavelength error per exposure is 15 km s$^{-1}$ for the G130M and G160M gratings and 150 km s$^{-1}$ for the G140L grating. Empirically, the observed wavelength positions of the detected galactic interstellar medium (ISM) lines (including \cii\ \ly 1334.53, \feii\ \ly 1608.45, and \alii\ \ly 1670.79) are in agreement between the two epochs to within 0.5\AA\ ($\sim$ 120 km s$^{-1}$). For the G130M and G160M gratings, the detected ISM line positions are also consistent with the laboratory values within 0.03\AA\ ($\sim$ 7 km s$^{-1}$). These wavelength errors are much smaller than the observed velocity shift (--1550 km s$^{-1}$) between the 2011 and 2017 epochs described in section \ref{section:acc}.

Detailed information about the observations and data reduction is given in section 2 of Paper II, where a total of four outflow systems (S1 -- S4, see table \ref{tb:OutflowSystems} here) were identified. The troughs from S1 show double-minima features, which are most apparent in the \naix, \arvii, and \arviii\ troughs (see figure 3 of Paper II). Since these two features appear at the same velocity in several troughs, S1 is divided into two components, 1a and 1b. The four lower velocity systems (S1a, S1b, S2, and S3) are consistent with no variations between the 2011 and 2017 epochs, and we report their analysis in Paper II. Here, we focus on the acceleration and physical characteristics of S4.

\begin{deluxetable}{ c c c}[htb!]
\tablewidth{0.5\textwidth}
\tabletypesize{\small}
\setlength{\tabcolsep}{0.02in}
\tablecaption{Outflows Detected in the SDSS J1042+1646 Data}
\tablehead{
 \colhead{Outflow \Comp}	& \colhead{Velocity$^{a}$ }	& \colhead{ \neviii\ Abs. Width$^{b}$} 			
\\
\\ [-2mm]
 \colhead{}  & \colhead{(km s$^{-1}$)} 	& \colhead{(km s$^{-1}$)}		
}

\startdata
\hline
S1a			&	\textbf{-4950} 		&	1700			\\
S1b			&	\textbf{-5750} 		&	1700			\\
S2			&	\textbf{-7500} 		&	1500			\\
S3			&	\textbf{-9940} 		&	1350			\\
S4, 2011		&	\textbf{-19500} 	&	2000			\\
S4, 2017		&	\textbf{-21050} 	&	2000			\\

\vspace{-2.2mm}
\enddata

\tablecomments{
\\
$^{a}$ The velocity centroids come from the Gaussian profile fitting to unblended absorption troughs, e.g., \arviii\ \ly\ly 700.24, 713.80.\\
$^{b}$ \neviii\ \ly 770.41 absorption trough width is measured for continuous absorption below a normalized flux of $I$ = 0.9. \\
}
\label{tb:OutflowSystems}
\end{deluxetable}

\section{Evidence for Outflow Velocity Shift}
\label{section:acc}

\begin{figure}[htp]

\centering
	\includegraphics[angle=0,trim={0cm 0.2cm 10.5cm 7cm},clip=true,width=1\linewidth,keepaspectratio]{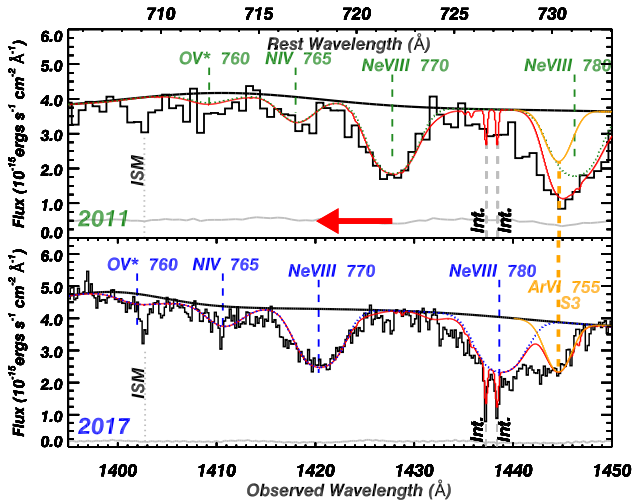}

\caption{Velocity shift of the outflow S4 in SDSS J1042+1646. The top and bottom panels are for the 2011 and 2017 epochs, respectively. The data are shown as black histograms while the errors are shown as solid gray lines. We show the Gaussian fitting for the strong ionic absorption troughs in green and blue dotted lines for the 2011 and 2017 epochs, respectively (see details in section \ref{sec:iden}). The absorption trough of \arvi\ \ly 754.93 from S3 is shown as the orange line in both panels (S3 does not vary between the two epochs). The combined absorption model in each panel is made by summing up all components and is shown as a solid red line. A strong Galactic ISM line (\siiv\ \ly 1402.77) and intervening systems are marked by gray lines. The red arrow shows the direction and magnitude of the outflow shift from the 2011 to 2017 epoch.\\}

\label{fig:spec1}
\end{figure}

\begin{figure}[htp]

\centering
	\includegraphics[angle=0,trim={0.0cm 0.0cm 15.3cm 0.5cm},clip,width=1.0\linewidth,keepaspectratio]{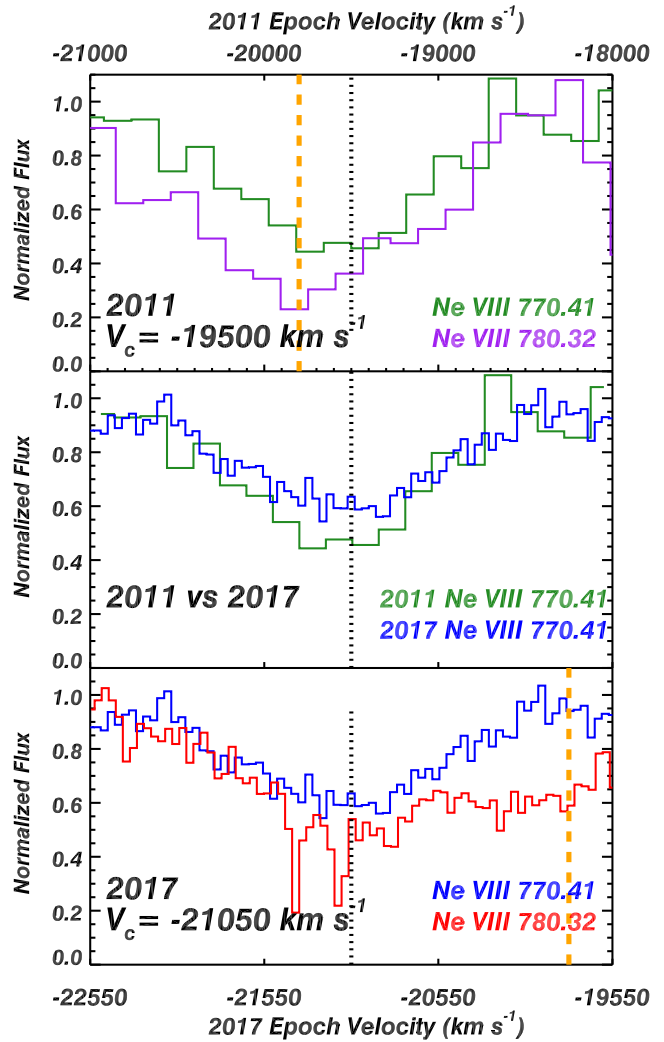} 
  
\caption{Comparison of the observed \neviii\ doublet absorption troughs between the 2011 and 2017 epochs. The velocity centroids ($v_c$) for each epoch are marked by the vertical, black dotted lines. The top and bottom panels' $x$-axes have a difference of -1550 km s$^{-1}$ in order to align $v_c$ for the 2011 and 2017 epochs. In the middle panel, we compare the \neviii\ \ly 770.41 between the two epochs, where the 2011 trough is shifted by --1550 km s$^{-1}$ (using the bottom velocity x-axis). The orange dashed lines point to the absorption troughs of \arvi\ \ly 754.93 from outflow S3 (see section \ref{sec:iden}). The narrow intervening absorption systems seen in the 2017 observation in the \neviii\ \ly 780.32 (at $\sim$ --21,100 km s$^{-1}$ and 21,300 km s$^{-1}$) are out of the velocity range in the top panel, which covers the 728 -- 733\angstrom\ rest-frame region (see figure \ref{fig:spec1}).}

\label{fig:NeVIIIcompare}
\end{figure}

\begin{figure}[htp]

\centering
	\includegraphics[angle=0,trim={0cm 0.2cm 10.5cm 7cm},clip=true,width=1\linewidth,keepaspectratio]{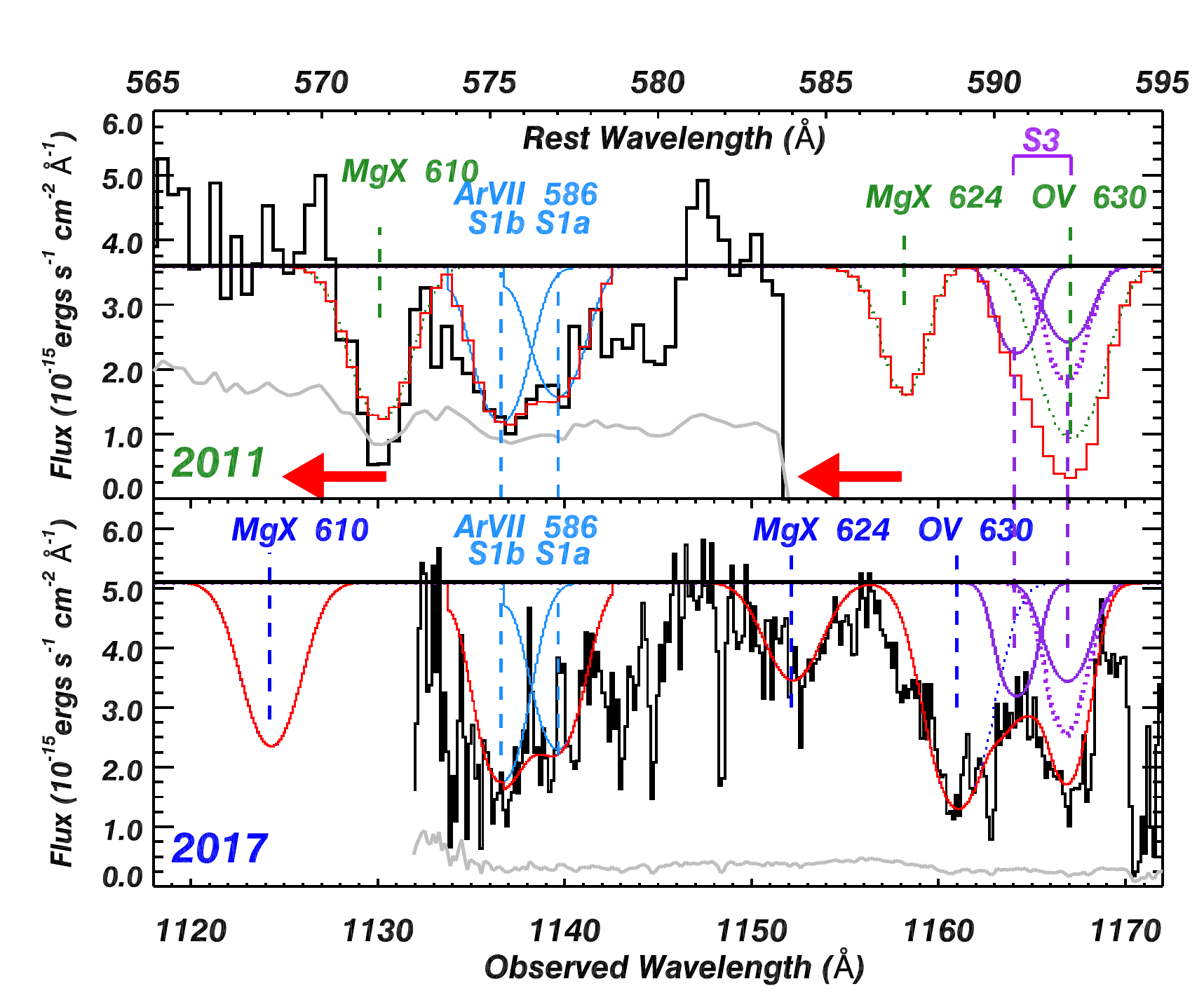}

\caption{Spectral region for \mgx\ \ly\ly 609.79, 624.94 and \ov\ \ly 629.73 of S4 in SDSS J1042+1646. The labels and colors for S4 are the same as in figure \ref{fig:spec1}. The absorption troughs of \arvii\ \ly 585.75 from S1a and S1b are shown as the light-blue lines. The absorption troughs from S3 are shown as the purple solid lines (\oiv\ \ly 608.40 and \oiv*\ \ly 609.83) and purple dotted line (\mgx\ \ly 609.79). S1a, S1b, and S3 did not vary between the two epochs. The red arrows show the direction and magnitude of the outflow shift from the 2011 to 2017 epoch. For S4, we identify the \mgx\ \ly 609.79 trough of the 2011 epoch as well as the \mgx\ \ly 624.94 and \ov\ \ly 629.73 troughs for the 2017 epoch.  See the discussion in section \ref{sec:MgX}.\\}
\label{fig:spec2}
\end{figure}

\begin{deluxetable}{c c l c }[htb!]
\tablewidth{0.47\textwidth}
\tabletypesize{\small}
\setlength{\tabcolsep}{0.02in}
\tablecaption{Column Densities for Outflow S4 in SDSS J1042+1646}
\tablehead{
	 \colhead{Ion}		& \colhead{ $\lambda$$^{(1)}$}		& \colhead{ N$_{ion,mea}$$^{(2)}$} 		 & \colhead{ $\frac{\text{N}_{ion,mea}}{\text{N}_{ion,model}}$$^{(3)}$}  			
\\
\\ [-2mm]
	 \colhead{} 		& \colhead{(\angstrom)}		& \colhead{log(cm$^{-2}$)}	 		& \colhead{}		
}

\startdata

\multicolumn{4}{l}{\textbf{Outflow \Comp\ 4, 2011 epoch, v = [-20800,-18600]$^{(4)}$}}\\ 
\hline
			\hi		&1025.72		&\color{red}$<$15.23			&$<$3.39		\\
			\niv		&765.15			&\color{red}$<$14.29			&$<$29.5		\\
			\ov	 	&629.73			&--$^{(5)}$				&--			\\
			\ovi	 	&1031.91		&\color{blu}$>$16.08			&$>$1.00		\\
			\neviii		&770.41, 780.32		&\color{blu}$>$15.98			&$>$0.56		\\
			\naix	 	&694.15$^{(5)}$		&\color{red}$<$15.30			&$<$6.3		\\
			\mgx	 	&609.79			&\color{blu}$>$15.73			&$>$1.00			\\
			\arviii		&700.24, 713.80		&\color{red}$<$14.46			&$<$60.3		\\
				 		
\hline
\multicolumn{4}{l}{\textbf{Outflow \Comp\ 4, 2017 epoch, v = [-22150,-20000]$^{(4)}$}}\\ 
\hline
			\hi		&949.74			&\color{red}$<$16.10			&$<$25.1		\\
			\niv		&765.15			&\color{red}$<$14.20			&$<$24.5		\\
			\oiv		&787.71			&\color{red}$<$14.59			&$<$13.8		\\
			\ov	 	&629.73			&\color{blu}$>$15.16			&$>$1.00		\\
			\ovi	 	&1031.91,1037.62	&--$^{(6)}$					&--		\\
			\neviii		&770.41, 780.32		&\color{blu}$>$16.06			&$>$0.68		\\
			\naix	 	&682.72			&\color{red}$<$15.30			&$<$6.16		\\
			\mgx	 	&624.94	$^{(6)}$	&\color{blu}$>$15.73			&$>$1.00		\\
			\arviii		&700.24			&\color{red}$<$14.56			&$<$75.8		\\
\vspace{-2.2mm}

\enddata

\tablecomments{
\\
$^{1}$ The rest wavelengths for the measured transitions. For doublet or multiplet transitions, we only show \ly\ for measured troughs. \\
$^{2}$ The measured column density (N$_{ion}$) for each ionic transition (see section \ref{sec:SSS}). Lower limits are shown in {\color{blu}blue}, while upper limits are shown in {\color{red}red}.\\
$^{3}$ The measured N$_{ion}$ divided by the model predicted N$_{ion}$.\\
$^{4}$ The N$_{ion}$ integration range in km s$^{-1}$.\\
$^{5}$ \ov\ \ly 629.73, \naix\ \ly 681.72, and \mgx\ \ly 624.94 fall into the gap of the COS G140L grating for the 2011 epoch (see figure \ref{fig:spec2}). \\
$^{6}$ \ovi\ \ly\ly 1031.91,1037.62 and \mgx\ \ly 609.79 are out of the observation range of the COS G130M grating for the 2017 epoch (see figure \ref{fig:spec2}).
}

\label{tb:IonSystems2}
\end{deluxetable}

\subsection{The \neviii\ Troughs}
\label{sec:iden}
We present the 1395 -- 1450\angstrom\ observed frame region in figure \ref{fig:spec1}, where the top and bottom panels are for the 2011 and 2017 epochs, respectively. The data are shown in black histograms, while the corresponding errors are shown as the gray lines. The orange lines are the models for the \arvi\ \ly 754.93 absorption trough from outflow S3, which is a stable outflow with no observed variability (see Paper II). We observe deep absorption troughs from the \neviii\ doublet at 770.41\angstrom\ and 780.32\angstrom\ in both epochs. Since their optical depth ratios are close to unity, the troughs are saturated. The \neviii\ absorption trough widths are 2000 km s$^{-1}$ for both the 2011 and 2017 epochs (see table \ref{tb:OutflowSystems}). Therefore, according to the BAL definition for the EUV500 band discussed in Paper I, S4 is classified as a BAL outflow.

There is an apparent wavelength shift of the troughs for the \neviii\ doublet between the two epochs, as indicated by the red arrow. In order to identify this observed shift, we first fit the \neviii\ troughs in the 2017 epoch with Gaussian optical depth profiles. We fix their Gaussian velocity centroids at --21,050 km s$^{-1}$, widths at $\sigma$ = 360 km s$^{-1}$, and depth ratio at 1:1. We scale their depths until the models fit the observed absorptions near 1420\angstrom\ and 1438\angstrom\ in the observed frame. These Gaussians are shown as the blue dotted lines on the bottom panel of figure \ref{fig:spec1}, and they fit the 2017 epoch's data well. We then apply the same Gaussian profiles (i.e., the same widths and depths) to the 2011 epoch but shift the velocity centroid to --19,500 km s$^{-1}$. These Gaussians are shown as the green dotted lines on the top panel of figure \ref{fig:spec1} and they fit well the troughs seen at 1428\angstrom\ and 1445\angstrom\ (observed frame). Both absorption features from the \neviii\ doublet have the same velocity shift ($\Delta v$) between the epochs and are well fitted with the same Gaussian width. These kinematic coincidences strongly suggest that we see the same outflow, but it has shifted by 1550 km s$^{-1}$ during the six-yr interval between the two epochs (3.2 yr in the quasar rest frame).



\subsection{Velocity Profiles Comparisons}
\label{sec:iden2}
To compare the velocity structure of the two epochs, figure \ref{fig:NeVIIIcompare} shows the \neviii\ absorption troughs in velocity space. The top and bottom panels are for the 2011 and 2017 epochs, respectively. The $x$-axes of the 2011 and 2017 epochs are shown at the top and bottom, respectively, and are shifted by --1550 km s$^{-1}$. The blending from the \arvi\ absorption trough of outflow S3 stays at the same velocity, and we mark the \arvi\ velocity centroids with orange dashed lines. The velocity centroids of the \neviii\ doublet for the two epochs are $v_{c,2011}$ = --19,500 km s$^{-1}$ and $v_{c,2017}$ = --21,050 km s$^{-1}$ (marked with the black dotted lines). 

In the top panel, the lower-velocity wings of the \neviii\ doublet troughs are similar, while the higher-velocity portion of the \neviii\ \ly 780.32 is contaminated by the \arvi\ \ly 754.93 absorption trough of S3. Similarly, in the bottom panel, the higher-velocity wings of the \neviii\ doublet are nearly identical, while the lower-velocity portion of the \neviii\ \ly 780.32 is contaminated by the stationary \arvi\ \ly 754.93 absorption trough of S3. These matches support the idea that the troughs we observed in both epochs come from the \neviii\ doublet transitions, which only partially cover the source and show non-black saturation.

In the middle panel of figure \ref{fig:NeVIIIcompare}, we compare the \neviii\ \ly 770.41 absorption troughs between the two epochs, and it is evident that the two epochs' \neviii\ \ly 770.41 troughs have nearly identical velocity structures when shifted by 1550 km s$^{-1}$. This strengthens the claim that the \neviii\ absorption troughs indeed shifted in velocity between 2011 and 2017 while the velocity profile remained unchanged.


\subsection{Support from the \mgx\ and \ov\ Troughs}
\label{sec:MgX}
\begin{figure}[htp]

\centering
	\includegraphics[angle=0,trim={0cm 2.46cm 0.3cm 1.5cm},clip=true,width=1\linewidth,keepaspectratio]{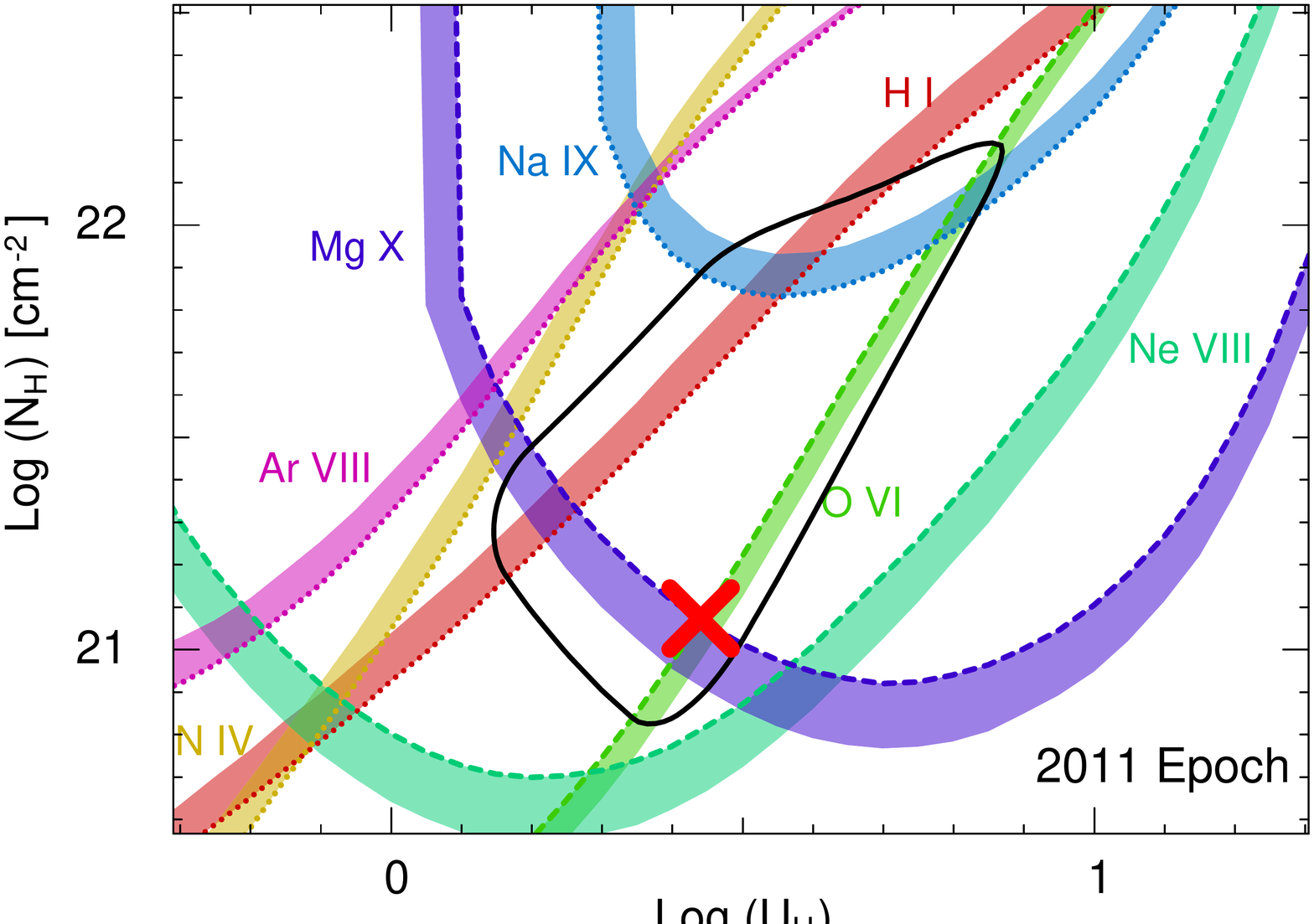} 
	\includegraphics[angle=0,trim={0cm 0cm 0.3cm 1.7cm},clip=true,width=1\linewidth,keepaspectratio]{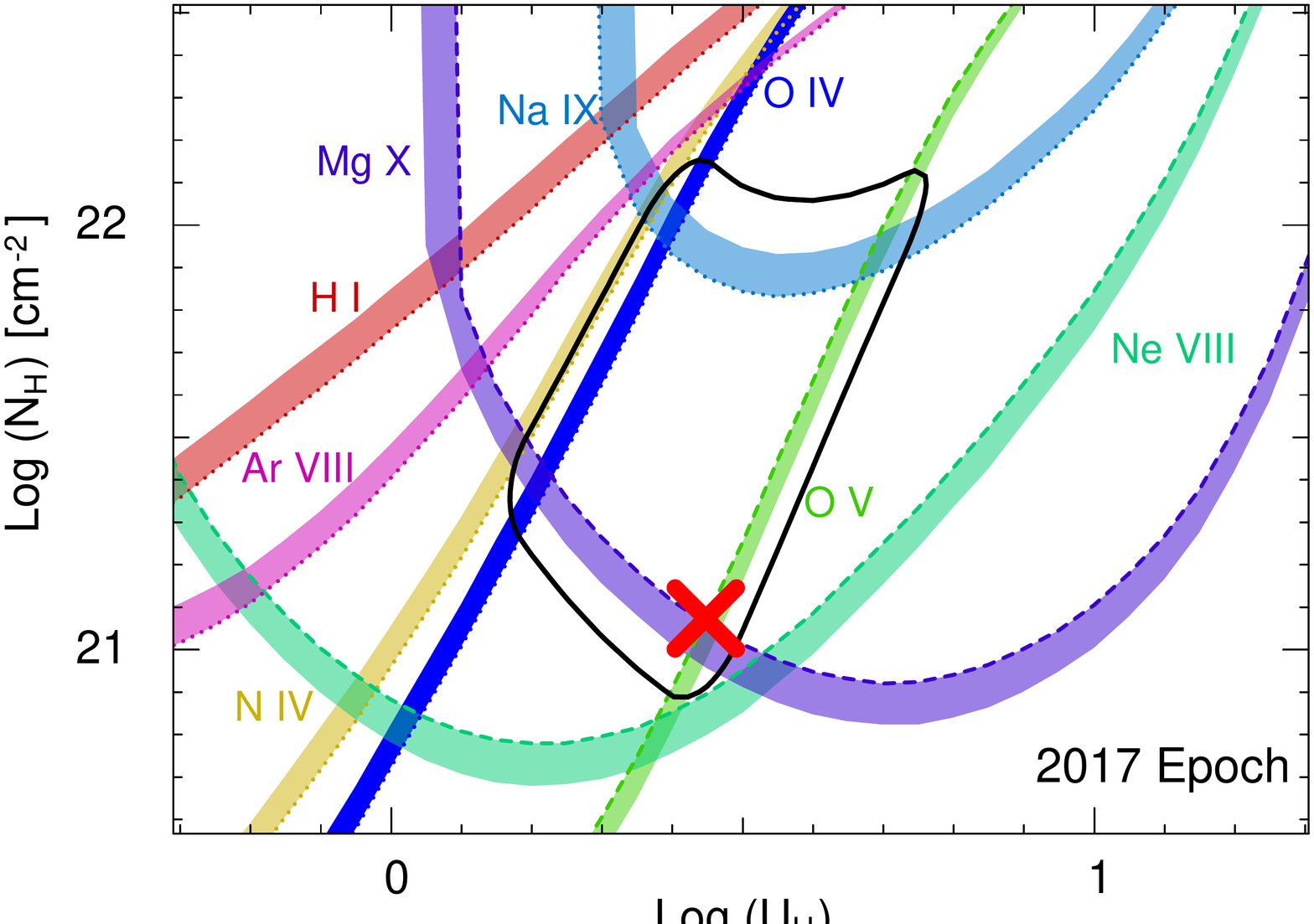} 

\caption{Photoionization solutions of outflow S4 for the 2011 and 2017 epochs. Each colored contour represents the region of \Nh\ and \Uh\ where the model predicts consistent N$_{ion}$ with the observed ones within the errors. The dashed lines represent \Nion\ lower limits, which allow the phase spaces above the lines. The dotted lines represent \Nion\ upper limits, which allow the phase spaces below the lines. For each panel, the region within the black line is the allowed photoionization solution bounded by a 1$\sigma$ error contour. The red crosses mark the adopted photoionization solutions for figure \ref{fig:spec2}.}
\label{fig:comp4}
\end{figure}

The Synthetic Spectral Simulation (\SSS) method creates a modeled spectrum based on the photoionization solution of the outflow (see section 3.3 of Paper II and section \ref{sec:SSS} here). In figure \ref{fig:spec2}, by using the photonionization solution derived in section \ref{sec:SSS} (the red crosses in figure \ref{fig:comp4}), we show the spectral region from about 1118\angstrom\ to 1172\angstrom\ (the observed frame), where we expect to observe the absorption troughs of \mgx\ \ly\ly 609.79, 624.94 and \ov\ \ly 629.73 from S4. The data and corresponding errors are shown as the black and gray histograms, respectively. Using Equations (2) and (3) from Paper II, we indicate the expected \SSS\ model centroids of the absorption troughs for the 2011 and 2017 epochs in green and blue dashed lines, respectively. The absorption troughs of \arvii\ \ly 585.75 from outflow S1a and S1b do not vary between the two epochs, and we show both of them as the light-blue lines. Similarly for S3, the absorption troughs remain at the same velocity between the two epochs. We show them as purple solid lines (\oiv\ \ly 608.40, \oiv*\ \ly 609.83) and a purple dotted line (\mgx\ \ly 609.79). For the 2011 epoch, the absorption troughs from \ov\ \ly 629.73 and \mgx\ \ly 624.94 fall into the gap of the COS G140L grating (1152 -- 1185\angstrom\ in the observed frame). The velocity centroid and width of the \SSS\ model for the \mgx\ \ly 609.79 trough are fixed. Therefore, the good fit of this modeled absorption to the trough seen at 1130\angstrom\ identifies the latter as the \mgx\ \ly 609.79 trough of S4 in the 2011 epoch. Our 2017 model predicts a --1550 km s$^{-1}$ shifted \mgx\ \ly 624.94 trough that matches well with the absorption trough seen at 1152\angstrom\ (the observed frame). \mgx\ \ly 609.79 and \ly 624.94 are doublet transitions that arise from the same ion, and the wavelength separation between them is fixed. Therefore, observing similar troughs for \mgx\ \ly 609.79 in the 2011 epoch and \mgx\ \ly 624.94 in the 2017 epoch, shifted by --1550 km s$^{-1}$, is strong evidence that the outflow shifted in velocity over the six years.






Finally, the \SSS\ model predicts an \ov\ \ly 629.73 trough consistent with the observed absorption near 1160\angstrom\ in the 2017 epoch. The combined absorption models (made by summing up all components) are shown as solid red lines in figure \ref{fig:spec2}.

\subsection{Summary of Outflow Velocity Shift Evidence}
The evidence for the velocity shift exhibited by outflow S4 is summarized as follows:

\indent 1. We identified in each epoch the \neviii\ \ly\ly 770.41, 780.32 doublet troughs where the 2017 epoch's troughs are shifted by --1550 km s$^{-1}$ (section \ref{sec:iden}).\\
\indent 2. The kinematic similarity of these \neviii\ troughs secures their identification as arising from the \neviii\ doublet (section \ref{sec:iden2}).\\
\indent 3. The existence of troughs at the expected wavelength and velocity width for the \mgx\ \ly 609.79 absorption in the 2011 epoch and \mgx\ \ly 624.94 absorption in the 2017 epoch (--1550 km s$^{-1}$ shifted compared to the 2011 epoch, section \ref{sec:MgX}).\\
\indent 4. The existence of a trough with the expected wavelength and shape of the predicted \ov\ \ly 629.73 trough in the 2017 epoch (section \ref{sec:MgX}).\\


\section{Photoionization Analysis}
\subsection{Ionization Solution}
\label{sec:SSS}
The ionic column densities (\Nion) we measure are representative of the ionization structure for the outflowing material. With the aid of photoionization models, we can determine the physical characteristics of the outflow. We follow the \SSS\ method in Paper II to derive the best fitting photoionization solution for both epochs separately. Here, we give a concise description of the \SSS\ method (see the full discussion in section 3.1 of Paper II).

1. We first measure the column densities (\Nion) from uncontaminated absorption troughs. Since there are no measurable doublet transitions in S4, we measure the \Nion\ of absorption troughs using the apparent optical depth method \citep[AOD; see e.g.,][]{Savage91}. Adopting the same criteria as in Paper II, for singlet transitions with a maximum optical depth, $\tau_{max}$, greater than 0.5, we treat the AOD \Nion\ as lower limits. For absorption troughs with $\tau_{max}$ $<$ 0.05, we take the AOD \Nion\ as upper limits. None of the uncontaminated troughs have 0.05 $<$ $\tau_{max}$ $<$ 0.5. The measured \Nion\ are shown in table \ref{tb:IonSystems2}.

2. A photoionization solution (PI$_{1}$) is built based on these measured \Nion. Photoionized plasma in a quasar outflow is characterized by the total hydrogen column density, \Nh, and the ionization parameter, \Uh, where
\begin{equation}
\label{Eq:ionPoten}
\Uh=\frac{\Qh}{4\pi R^2 \nh c}
\end{equation}
where $\Qh$ is the source emission rate of hydrogen ionizing photons, $R$ is the distance of the outflow from the central engine, $\nh$\ is the hydrogen number density (for a highly ionized plasma, \ne\ $\simeq$ 1.2 \nh), and c is the speed of light.

3. We run the spectral synthesis code Cloudy [version c17.00, \cite{Ferland17}] to generate grids of photoionization simulations. At each grid point, Cloudy predicts the N$_{ion}$ for all ions in its database.

We assume a solar metallicity and adopt the spectral energy distribution (SED) of HE 0238 SED, which is based on the EUV500 observations of quasar HE 0238--1904 \citep{Arav13}. We use the HE 0238 SED since: a) for the observed data (570--1000 \AA\ rest frame), the ratio of the HE 0238 SED with respect to the SDSS J1042+1646 continuum is constant to within $\pm$ 10\%; b) The observation of quasar HE 0238--1904 has higher  signal to noise; and c) It allows us to compare the physical parameters of outflows from different objects with the same baseline SED. By integrating this SED, the bolometric luminosity of SDSS J1042+1646 is $\sim$ 1.5 $\times$ 10$^{47}$ erg s$^{-1}$.

We present the derived photoionization solutions for both epochs in figure \ref{fig:comp4}. The dashed lines represent \Nion\ lower limits, which allow the phase spaces above the lines. The dotted lines represent \Nion\ upper limits, which allow the phase spaces below the lines. For each ion, we add the measured \Nion\ error with an additional 20\% error in quadrature (accounting for systematic errors, see section 3.1 of Paper III), and treat this combined value as the final error. Since all \Nion\ are lower or upper limits, large regions in these phase spaces contain acceptable solutions, i.e., within the black contours. Both epochs show consistent solutions with log(\Nh) between 20.8 and 22.2 [hereafter, \Nh\ is in units of log(cm$^{-2}$)] and log(\Uh) between 0.2 and 0.9. This consistency supports the claim that the outflow we observed in 2011 and 2017 are the same outflow, shifted by 1550$^{+150}_{-150}$  km s$^{-1}$.

We also explored other SEDs [MF87 \citep{Mathews87}, ultraviolet (UV)-soft \citep{Dunn10}] and metallicities (super solar, Z = 4.67Z$_{\odot}$, described in section 3.2 of Paper V). These choices change the log(\Uh) of the solution by less than 0.3 dex and lower log(\Nh) by up to 0.6 dex. The analysis of the velocity shift is not affected by these \Nh\ and \Uh\ differences.

4. We assume that all troughs in S4 can be modeled with similar Gaussian profiles (i.e., the same velocity centroid and width, see equations (2) and (3) in Paper II). By adopting the predicted \Nion\ from Cloudy, we create an AOD synthetic spectrum model for the entire observed spectrum. In figure \ref{fig:spec2}, we show this synthetic spectrum using the photoionization solutions marked by the red crosses in figure \ref{fig:comp4}.


\subsection{Determination of \ne\ from the \ov*\ Multiplet}
\label{sec:ne}
\ov*\ has a multiplet of six transitions that create absorption troughs near 760\angstrom\ (the rest frame), and they are sensitive to a wide range of electron number densities (\ne) (see section 4.2.3 in Paper II). We detect absorption at the expected wavelength locations of the \ov*\ multiplet. We adopt the same analysis from Paper II as follows. Since the derived photoionization solutions allow large regions in the phase space (black contours in figure \ref{fig:spec2}), we check the photoionization solutions on the boundary of the contours and constrain \ne.

\begin{figure}[htp]

\centering
	\includegraphics[angle=0,trim={0cm 7.5cm 0cm 4cm},clip=true,width=1\linewidth,keepaspectratio]{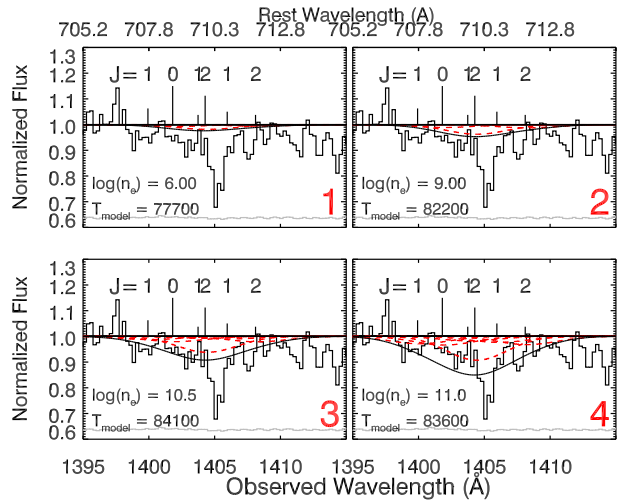}\\
\caption{Fits to the \ov*\ multiplet region for outflow S4. We vary \ne\ (in units of cm$^{-3}$) to get the best fit. The \ne\ and the corresponding temperature predicted from Cloudy are shown at the bottom-left corner of each panel. The black and gray solid histograms are the normalized flux and errors for the 2017 epoch. For each subplot, the red dashed lines represent the models of the \ov*\ multiplet for a particular log(\ne) , while the solid black lines are the summation of all models in this region. We start the y-axis from 0.6 to highlight the shallow \ov*\ troughs, and we added 0.6 to the errors correspondingly. See section \ref{sec:ne} for a detailed discussion.}
\label{fig:ne_comp4}
\end{figure}

For the 207 epoch, we start with using the photoionization solution marked as the red $\times$ sign in the bottom panel of figure \ref{fig:spec2}, we adopt the model predicted value of N(\ov) and the temperature. We vary log(\ne) from 4 to 12 [hereafter, log(\ne) is in units of cm$^{-3}$] and overlay the model predicted \ov*\ troughs to the 1395\angstrom\ -- 1415\angstrom\ observed frame region (see figure \ref{fig:ne_comp4}). The red dashed lines represent the models of the \ov*\ multiplet for a particular \ne, and the solid black lines are the summation of all models in this region.  The model with log(\ne)~=~6 predicts minimal absorption troughs and clearly underestimates the observed trough, while the model with log(\ne)~=~11 overestimates the observations by more than 1$\sigma$. The absorption troughs are fitted well by the models with log(\ne) between 9 to 10.5 for the 2017 epoch. We then do similar analysis and constrain \ne\ adopting different photoionization solutions on the boundary of the black contours in figure \ref{fig:spec2}. Combining all \ne\ constraints, we get 4.5 $<$ log(\ne) $<$ 10.5. Incorporating this range with the derived photoionization solution, we constrain the distance ($R$) of this outflow in the range of 0.05 pc $<$ $R$ $<$ 54.3 pc. 

The thickness of outflow S4 is $\Delta R$ $=$ \Nh/\ne\ $<$ 0.01 pc. Therefore, the assumption that $\Delta R$ $\ll$ $R$ is valid and we can use equations (6) and (7) in \cite{Borguet12a} to calculate the mass flow rate ($\dot{M}$) and kinetic luminosity ($\dot{E}_{k}$) of the outflow. The derived $R$ values leads to a range of 0.07 $M_{\odot}$ yr$^{-1}$ $<$ $\dot{M}$ $<$ 141.5 $M_{\odot}$ yr$^{-1}$\ and 1.0 $\times$ 10$^{43}$ erg s$^{-1}$ $<$ $\dot{E}_{k}$ $<$ 2.0 $\times$ 10$^{46}$ erg s$^{-1}$. For the 2011 epoch, the signal to noise and spectral resolution are lower, but the absorption troughs in the \ov*\ region are consistent with the 2017 ones (see figure \ref{fig:spec1}). Caveat: This analysis is based on the assumption that most of the observed absorption in this region is from the \ov*\ multiplet. If the observed absorption is not from \ov*, we have the above derived \ne\ as an upper limit, log(\ne) $<$ 10.5 and $R$ as a lower limit, $R$ $>$ 0.05 pc. \\

\section{Discussion}
\subsection{Excluding Other Explanations For The \\Shifted Troughs}
\label{sec:BALacc}

The S4 outflow is a good candidate for an accelerating BAL outflow since the two epochs show not only similar velocity profiles but also close photoionization solutions. However, several other causes may explain the observed shift of the troughs:\\

1)  Transverse motion of the outflowing material across the LOS \citep[e.g.,][]{Moe09, Capellupo12,Yi19}. If this is the case, the BAL disappearance (the cloud moving out of the LOS) at -19,500 km s$^{-1}$ and a new BAL appearance (the cloud moving into the LOS) at -21,050 km s$^{-1}$ need to happen during the same 3.2 years interval (the quasar rest-frame time). This is improbable since individually observed BAL appearance and disappearance rates are low \citep[2.3 -- 3.9 \%, ][]{Filiz12, DeCicco17, McGraw17}, where their observations span 0.3 -- 4.9 years rest-frame timescales. \\

2) Instrumental artifacts. We note that the 2011 epoch has a short exposure time ($\sim$ 900s). In order to exclude any possible instrumental artifacts, we carefully looked at the data quality flags and calibration of the other observations taken close to our 2011 epoch observation. We found no possible instrumental issues which could significantly affect the region of interest (1395 -- 1450\angstrom\ in the observed frame). We also checked the wavelength calibrations of both epochs as shown in section \ref{sec:Data}.\\

\begin{figure}[htp]

\centering
	\includegraphics[angle=0,trim={0cm 2.48cm 0.3cm 1.5cm},clip=true,width=1\linewidth,keepaspectratio]{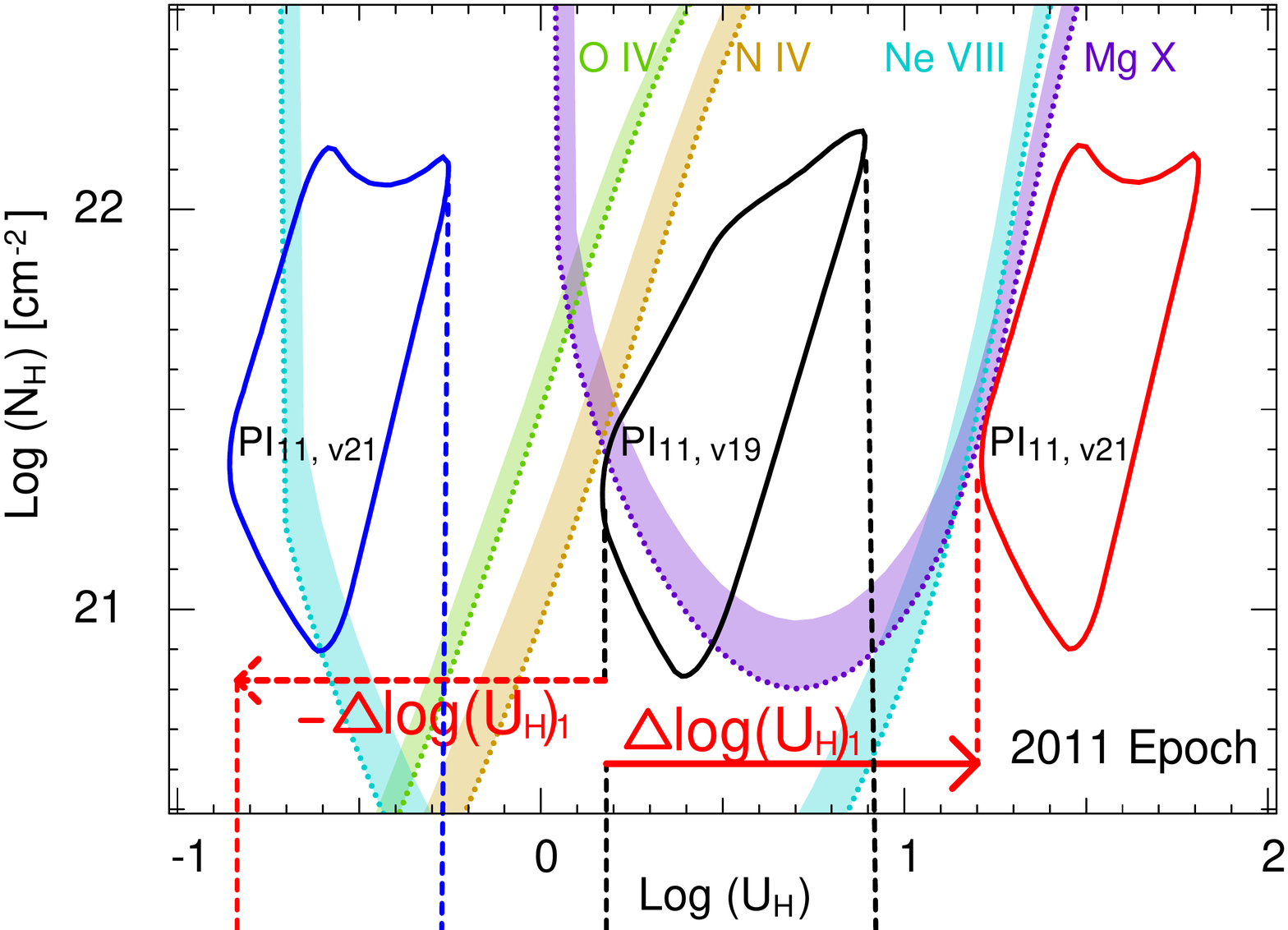} 
	\includegraphics[angle=0,trim={0cm 0cm 0.3cm 1.98cm},clip=true,width=1\linewidth,keepaspectratio]{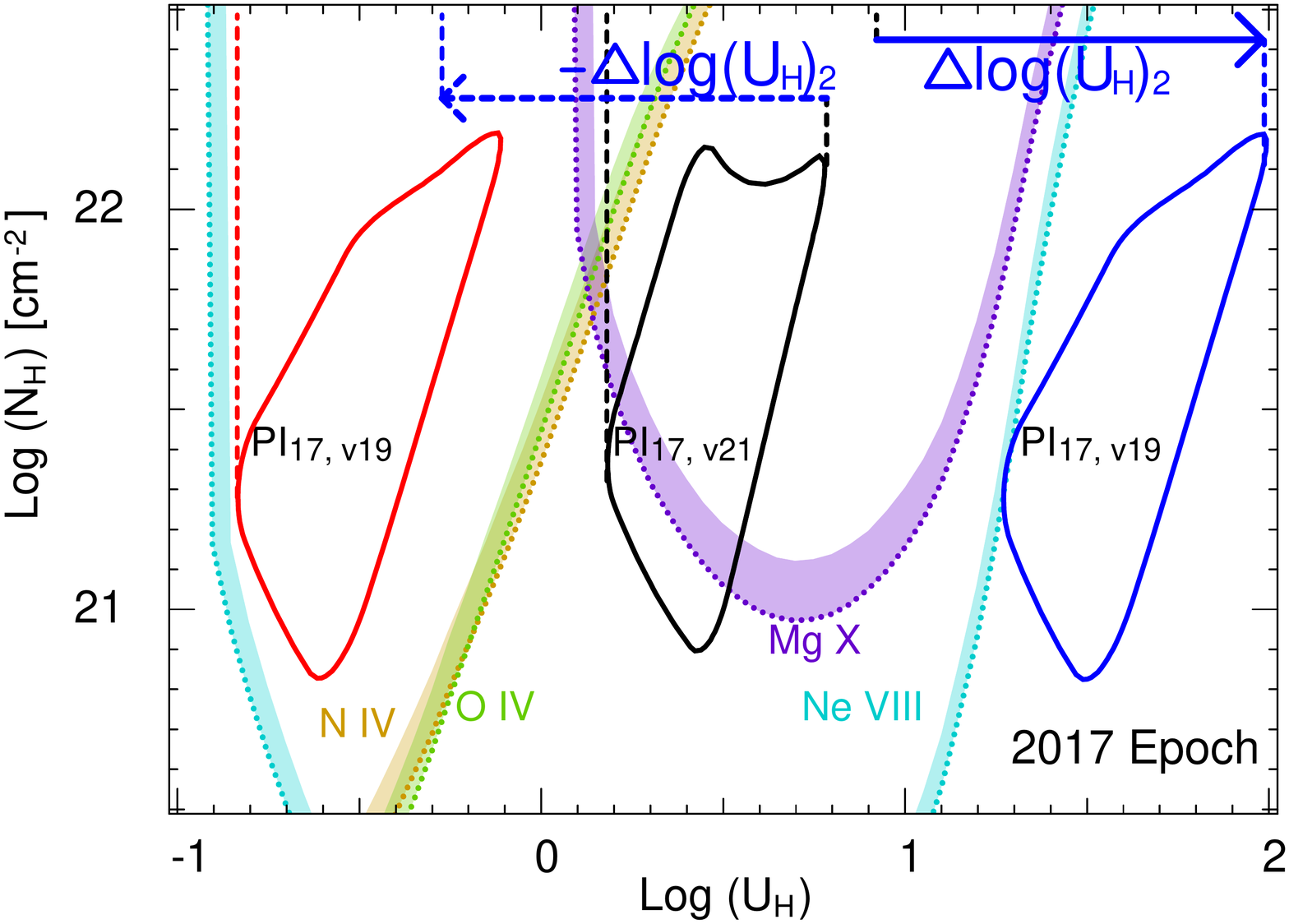} 
	
\caption{Phase plot for demonstrating the inability of a time-dependent photoionization model to explain the observed troughs of outflow S4. \textbf{Top:} For the 2011 epoch, we show the derived photoionization solution of the outflow centered at v$_{c}$ = -19,500 km s$^{-1}$ ($v19$, see section \ref{sec:SSS}) as the black contour and the measured \Nion\ upper limits for the v$_{c}$ = -21,050 km s$^{-1}$ ($v21$) outflow as the color dotted lines with their 1$\sigma$ error contours. \textbf{Bottom:} For the 2017 epoch, we show the derived photoionization solution of the outflow centered at v$_{c}$ = -21,050 km s$^{-1}$ as the black contour and the measured \Nion\ upper limits for the v$_{c}$ = -19,500 km s$^{-1}$ outflow as the color dotted lines. The minimal \Uh\ for a stationary $v21$ outflow in 2011 (red oval, top panel), predicts the red oval \Nh/\Uh\ solution for the $v19$ component in 2017. The latter is excluded by the measured column-density upper limits of \niv, \oiv\ and \neviii. Similarly, a high \Uh\ solution for the $v19$ outflow in 2017 (blue oval, bottom panel), is excluded by the predicted 2011 \Nh/\Uh\ solution of the $v21$ component (blue oval, upper panel). See elaboration in section \ref{sec:BALacc}. }
\label{fig:PhotoChange}
\end{figure}

3) Time-dependent photoionization changes. The idea is instead of having one outflow that accelerated from $v_c$ = -19,500 km s$^{-1}$ to -21,050 km s$^{-1}$ between the 2011 and 2017 epochs, there are 2 stationary outflows, one at each velocity. In this case, the changes in the absorption features between the two epochs are explained by changes in the incident ionizing flux of the quasar. 

This quasar (SDSS J1042+1646) has four other outflows (S1a, S1b, S2, and S3), which have absorption troughs consistent with no variability between the 2011 and 2017 epochs. These include doublet troughs that are clearly not saturated (e.g., the \arviii\ \ly\ly 700.24, 713.80 troughs in S2, see figure 6 of Paper II). Significant changes in \Uh\ would have caused large changes in these troughs, which are not detected.  Therefore, the large \Uh\ differences between the two epochs needed to explain the appearing and disappearing of the invoked S4 stationary outflows ($\Delta$log$(\Uh)$ $\simeq$ 1.0 dex, see quantitative analysis below) are excluded. 


We illustrate the possible photoionization change scenarios in figure \ref{fig:PhotoChange}. We denote the photoionization solutions as PI$_{\alpha,\beta}$, where $\alpha$ = 11 or 17 is for the 2011 and 2017 epochs, respectively, and $\beta$ = -$v19$ or -$v21$ corresponds to the outflows at $v_c$ = -19,500 km s$^{-1}$ and -21,050 km s$^{-1}$, respectively. In the top panel, we show the derived PI$_{11, v19}$ (see figure \ref{fig:comp4} and section \ref{sec:SSS}) as the black contour. Since we do not detect outflow troughs at $v$ = -21,050 km s$^{-1}$ in the 2011 epoch, we overlay in the top panel the measured \Nion\ upper limits for the $v21$ outflow from the 2011 epoch (colored dotted lines with corresponding 1$\sigma$ error contours). Similarly, in the bottom panel, we show the derived PI$_{17, v21}$ as the black contour and \Nion\ upper limits for the $v19$ outflow measured from the 2017 epoch as colored dotted lines (with corresponding 1$\sigma$ error contours). 

We assume that the \Nh\ for each outflow did not change significantly between the two epochs. Otherwise, we are in the regime of the transverse motion scenario, which we showed was improbable in point 1) above. We now ask the question: at what \Uh\ values would the $v21$ outflow be consistent with the \Nion\ upper limits of the 2011 epoch? To answer that, we take the (\Nh, \Uh) solution of the $v21$ outflow from the 2017 epoch (the black contour on the bottom panel of figure \ref{fig:PhotoChange}), and superimpose it on the top panel. To match the \Nion\ constraints, we are allowed to change only \Uh\ as \Nh\ is assumed constant. The minimal \Uh\ shift is shown by the position of the red contour in the top panel. Therefore, we can match the \Nion\ constraints for the $v21$ outflow from the 2011 epoch if \Uh\ of the $v21$ outflow from the 2017 epoch increases by at least $\Delta$log$(\Uh)_{1} $ = 1.0 dex (shown as the solid red arrow). 

Since the quasar is the only ionizing photon source, PI$_{17, v19}$ would have the same magnitude of \Uh\ shift with respect to PI$_{11, v19}$ but in the opposite direction ($-\Delta$log$(\Uh)_{1}$ and the dashed red arrow), and the corresponding PI$_{17, v19}$ is shown as the red contour in the bottom panel. We find that all (\Uh, \Nh) solutions within PI$_{17, v19}$ overestimate the measured \Nion\ upper limits of \niv\ and \oiv\ in the 2017 epoch by at least an order of magnitude. Thus, this is not a viable scenario.

Similarly, in the bottom panel, we do not detect outflow troughs from the $v19$ system in the 2017 epoch. We take the (\Nh, \Uh) solution of the $v19$ outflow from the 2011 epoch (the black contour on the top panel), and superimpose it on the bottom panel to match the measured \Nion\ upper limits here. This corresponds to a \Uh\ shift between PI$_{17, v19}$ (blue contour in the bottom panel) and PI$_{11, v19}$ (black contour in the top panel) of $\Delta$log$(\Uh)_{2} $ $>$ 1.1 dex, with the minimum shift marked as the solid blue arrow. PI$_{11, v21}$ would have the same magnitude of \Uh\ shift with respect to PI$_{17, v21}$ but in the opposite direction ($-\Delta$log$(\Uh)_{2} $ and the dashed blue arrow). Again, we find that all (\Uh, \Nh) solutions within PI$_{11, v21}$ (blue contour at the top panel) violate the measured \Nion\ upper limits of \niv\ and \oiv\ in the 2011 epoch by at least 1.1 dex, eliminating the validity of this scenario.

There are two additional concerns regarding the time-dependent photoionization, which need to be discussed. \\
a) Since $v19$ and $v21$ could be at different $R$ with respect to the central quasar, the inner outflow responds earlier to the change in ionizing flux than the outer outflow. However, since both outflows are observed in the LOS, the response of the inner outflow propagates towards us at the speed of light and coincides with the observed response of the outer outflow. Therefore, from our perspective, both outflows react to the change in the ionizing flux of the quasar simultaneously.\\
b) The reaction time for the outflows to reach photoionization equilibrium when the ionizing flux from the quasar changes. Both outflows have an electron number density (\ne) in the range of 10$^{4.5}$ -- 10$^{10.5}$ cm$^{-3}$ (see section \ref{sec:ne}). When \Uh\ decreases or increases by 1 dex, the response time for an ionic transition like \neviii\ in outflows with these \ne\ values is $\lesssim$ 10 days \citep[e.g.,][]{Kriss95}. From statistical studies of quasar variabilities \citep[e.g.,][]{Filiz13, DeCicco17}, BAL quasar outflows barely vary for rest-frame timescale of $<$ 1 year. Thus, both $v19$ and $v21$ outflows are likely in photoionization equilibrium when observed. These two points do not affect our exclusion of the photoionization-change scenario above.

Overall, we exclude the motion of material into and out of the LOS, instrumental artifacts, and time-dependent photoionization changes as alternate explanations. Therefore, the outflow acceleration scenario is the only viable physical cause for the observed --1550 km s$^{-1}$ velocity shift in S4.\\

\begin{deluxetable*}{ l l l l l}[htb!]
\tablewidth{0.88\textwidth}
\tabletypesize{\small}
\setlength{\tabcolsep}{0.02in}
\tablecaption{Comparisons of BAL Acceleration Candidates}
\tablehead{
 \colhead{References}	& \colhead{$\Delta t_{rest}^{a}$}	&\colhead{$\Delta v$}	& \colhead{ Accel.$^{b}$} & \colhead{ Accel.$^{b}$} 			
\\
\\ [-2mm]
 \colhead{}  		& \colhead{(year)} 		&\colhead{(km s$^{-1}$)}	& \colhead{(cm s$^{-2}$)} & \colhead{(km s$^{-1}$ yr$^{-1}$)}		
}

\startdata
\cite{Vilkoviski01}			&	\textbf{5.0} 		&\textbf{55}		&	\textbf{0.035$^{+0.016}_{-0.016}$}	&	\textbf{11$^{+5}_{-5}$}			\\	
\cite{Hall07}				&	\textbf{1.4} 		&\textbf{70}		&	\textbf{0.15$^{+0.025}_{-0.025}$}	&	\textbf{50$^{+8}_{-8}$}			\\
\cite{Grier16}	$^{c}$			&	\textbf{3.7} 		&\textbf{730}		&	\textbf{0.63$^{+0.14}_{-0.12}$}		&	\textbf{200$^{+44}_{-38}$}		\\
\cite{Grier16}	$^{c}$			&	\textbf{5.2} 		&\textbf{890}		&	\textbf{0.54$^{+0.04}_{-0.04}$}		&	\textbf{170$^{+13}_{-13}$}		\\
SDSS J1042+1646, S4 (This work)		&	\textbf{3.2} 		&\textbf{1550}		&	\textbf{1.52$^{+0.16}_{-0.16}$}		&	\textbf{480$^{+50}_{-50}$}		\\

\vspace{-2.2mm}
\enddata

\tablecomments{
\\
$^{a}$ The time intervals measured in the quasar rest-frame.\\
$^{b}$ The average acceleration measured in the quasar rest-frame.\\
$^{c}$ \cite{Grier16} reported two BAL acceleration candidates.
}
\label{tb:BALacc}
\end{deluxetable*}
\vspace{-0.2mm}

\begin{deluxetable}{ l l l l }[htb!]
\tablewidth{0.48\textwidth}
\tabletypesize{\small}
\setlength{\tabcolsep}{0.02in}
\tablecaption{Comparisons of Outflow Parameters to Quasar PG 1211+143}
\tablehead{
 \colhead{Outflow}	& \colhead{$v$}				&\colhead{log(\Nh)}		& \colhead{ log(\Uh)} 			
\\
\\ [-2mm]
 \colhead{}  		& \colhead{(km s$^{-1}$)} 		&\colhead{(log(cm$^{-2}$))}	& \colhead{} 
}

\startdata
SDSS J1042+1646, S4			&	\textbf{--21,000} 		&\textbf{20.8--22.2}		&	\textbf{0.2--0.9}			\\	
PG 1211+143				&	\textbf{--17,000} 		&\textbf{21.5}			&	\textbf{1.6}		\\

\vspace{-2.2mm}
\enddata


\label{tb:UFOcomp}
\end{deluxetable}
\vspace{-0.2mm}

\subsection{Comparisons with Other Studies}
\label{sec:Other}
As mentioned in section \ref{sec:Intro}, there are a few prior studies on BAL accelerations. In table \ref{tb:BALacc}, we summarize and compare them to our findings. We report the observed velocity shift in the third column, which is measured directly from the spectra. In the fourth and fifth columns, we report the average acceleration in the quasar's rest frame in two units, cm s$^{-2}$ and km s$^{-1}$ yr$^{-1}$. We note that the S4 outflow not only has the largest velocity shift observed  to date, i.e., 1550 km s$^{-1}$; but also has the largest BAL acceleration observed to date, i.e., 480$^{+50}_{-50}$ km s$^{-1}$ yr$^{-1}$ $=$ 1.52$^{+0.16}_{-0.16}$ cm s$^{-2}$ in the quasar rest-frame.

The previous studies of BAL acceleration in table \ref{tb:BALacc} detected a velocity shift only in the \civ\ \ly\ly 1548.19, 1550.77 absorption trough where, for BAL outflows, the absorption troughs from the \civ\ doublet usually blend together \citep[e.g.,][]{Grier16}. For S4, a consistent velocity shift signature is detected in four troughs, which is the first time that a quasar outflow velocity shift is observed from more than one ion and in distinct troughs from a doublet transition (\neviii, see section \ref{section:acc}).


\subsection{Similarity with the PG 1211+143 X-Ray Outflow}
\label{sec:UFO}
Outflow S4 has a similar velocity to the X-ray outflow seen in PG 1211+143 (--17,300 km s$^{-1}$). The latter is the only high-velocity outflow detected in X-ray grating spectra, which includes troughs from \nex-Ly$\alpha$, \mgxii-Ly$\alpha$, \sixiii-He$\alpha$ and \sixiv-Ly$\alpha$ using Chandra observations \citep[][]{Danehkar18}. \cite{Pounds16a,Pounds16b} also detected similar troughs using X-ray Multi Mirror (XMM-Newton) mission Reflection Grating Spectrometers (RGS) data. This X-ray absorber in PG 1211+143 is well fitted with log(\Nh) $\sim$ 21.5 and log($\xi$) $\sim$ 2.9, where $\xi$ is the X-ray ionization parameter. For the HE 0238 SED, we have the relation: log(\Uh) = log($\xi$) -- 1.3. The UV counterpart of this X-ray outflow has been detected in HST/COS observations, which yields a broad Ly$\alpha$ absorption feature at v = --17,000 km s$^{-1}$ (--0.056$c$) \citep{Kriss18}.  We compare the $v$, \Nh, and \Uh\ values between the X-ray outflow in PG 1211+143 and outflow S4 in table \ref{tb:UFOcomp}. 

We conclude that our observations in the EUV500 band have probed an outflow with similar velocity and \Nh\ to the one observed in PG 1211+143. The \Uh\ value of the PG 1211+143 [log(\Uh) = 1.6] is roughly an order of magnitude larger than what we find for S4, using the EUV500 data. This suggests that S4 may have an even higher-ionization phase similar to the one in PG 1211+143, which could be detected by future X-ray observatories \citep[e.g., \textit{Athena}, ][]{Barcons17}.

\subsection{BAL Acceleration and the Disk-wind Model}
\label{sec:DiskWind}
Radiatively accelerated disk-wind models \citep[e.g.,][]{Arav92, Murray95, Proga03, Proga04} are possible explanations for the origin of the observed BAL outflow. As shown in equation (7) of \cite{Murray95}, for a radiatively accelerated outflow, the radial velocity has the form:
\begin{equation}
    v(r) = v_\infty(1-r_{f}/r)^{\beta}
    \label{eq:vr}
\end{equation}
where $v(r)$ is the observed outflow velocity, $r_{f}$ is the launching radius of the outflow, $r$ is the outflow's current radius, and $\beta$ $\sim$ 1.15 \citep[full range 1.1 -- 1.2 from][]{Murray97}. 

The corresponding acceleration derived from equation (\ref{eq:vr}) is:
\begin{equation}
    a(r) \equiv \frac{dv}{dt} \equiv v\frac{dv}{dr} = 1.15\frac{v^2_\infty r_f}{r^2}\left (  1-\frac{r_f}{r}\right )^{1.30}
    \label{eq:ar}
\end{equation}

If the outflow is accelerated by radiation pressure, we have the terminal velocity of the outflow as:
\begin{equation}
v_\infty = F\sqrt{GM/r_{f}}
    \label{eq:vinf}
\end{equation}
where M is the central black hole's mass, G is the gravitational constant, and $F$ is scaling factor \citep[$\sim$ 1.5 -- 3.5, ][]{Murray95, Laor02, Baskin14}. Using equation (\ref{eq:vinf}), equations (\ref{eq:vr}) and (\ref{eq:ar}) can be rewritten as:


\begin{equation}
    v(r) = F\sqrt{GM/r_{f}}\left (1-\frac{r_{f}}{r}\right )^{1.15}
    \label{eq:vr2}
\end{equation}
and 
\begin{equation}
    a(r) = 1.15\frac{F^2GM}{r^2}\left (  1-\frac{r_f}{r}\right )^{1.30}
    \label{eq:ar2}
\end{equation}

\begin {table}[t]
\begin{center}

\caption{Predictions for BAL Accelerations in SDSS J1042+1646 (S4)}
\scalebox{0.94}{			
\setlength{\tabcolsep}{4pt}	
\begin{tabular}{llcc cc}

\hline  \\[-2.0ex]    
\multicolumn{1}{l}{Epoch} &
\multicolumn{1}{c}{$\Delta v^{a}$} &
\multicolumn{1}{c}{$v^{b}$} &
\multicolumn{1}{c}{$a^{c}$} &
\multicolumn{1}{c}{$a^{c}$} &
\multicolumn{1}{c}{$r^{d}$} 
\\

\multicolumn{1}{l}{} &
\multicolumn{1}{r}{(km s$^{-1}$)} &
\multicolumn{1}{r}{(km s$^{-1}$)} &
\multicolumn{1}{r}{(cm s$^{-2}$)} &
\multicolumn{1}{r}{(km s$^{-1}$ yr$^{-1}$)} &
\multicolumn{1}{r}{(pc)} 
\\
\hline \\[-1.8ex]

\textbf{2017$^{e}$} 	&\textbf{0}		&	\textbf{21050}	&	\textbf{1.52}		&	\textbf{480}		&	\textbf{0.23}		\\
\textbf{2019} 		&\textbf{400}		&	\textbf{21400}	&	\textbf{1.31}		&	\textbf{410}		&	\textbf{0.25}		\\	
\textbf{2022} 		&\textbf{800}		&	\textbf{21900}	&	\textbf{1.06}		&	\textbf{330}		&	\textbf{0.28}		\\
\textbf{2027} 		&\textbf{1500}		&	\textbf{22500}	&	\textbf{0.76}		&	\textbf{240}		&	\textbf{0.34}		\\

\hline \\[-5ex]

\label{tb:ACCpredict}
\end{tabular}}
\end{center}
\tablecomments{
\\
$^{a}$ The velocity difference of S4 between the predicted time and the 2017 epoch.\\
$^{b}$ Predicted outflow velocity of S4.\\
$^{c}$ Predicted accleration of S4 in the quasar's rest frame.\\
$^{d}$ Predicted outflow distance of S4 to the quasar.\\
$^{e}$ We show the measured parameters for the 2017 epoch as a comparison.
}
\end{table}

In our case, $v(r)$ = 21050 km s$^{-1}$ and $a(r)$ = 1.52 cm s$^{-2}$ are derived from the velocity shift between the 2011 and 2017 epochs (see section \ref{section:acc}). The Sloan Digital Sky Survey (SDSS) spectrum of J1042+1646 shows a \mgii\ broad emission line (BEL). By fitting this BEL using the \mgii--based black hole mass estimators \citep[see equation (7) and table (4) in][]{Bahk19}, we derived M $\sim$ 2.0$\times$10$^9$ $M_\odot$. The unknowns in equations (\ref{eq:vinf}), (\ref{eq:vr2}) and (\ref{eq:ar2}) are $F$, $r$, and $r_f$. As shown in section 4.1 of \cite{Grier16}, the disk-wind model of \cite{Murray97} will be viable if these equations are satisfied by the observations. To solve these equations, we vary $F$ between 1.5 and 3.5, and $r/r_f$ between 1 and 100, both in steps of 0.1. We find a good solution when $F$ = 1.8 and $r/r_f$= 5.7.  With these values of $F$ and $r/r_f$, the model predicts that the launching radius $r_{f}$ $\sim$ 1.2$\times$10$^{15}$ m (0.04 pc), the observing radius is 5.7 times $r_{f}$, i.e., $r$ $\sim$ 0.23 pc, and the terminal velocity, $v_{\infty}$ $\sim$ 26,000 km s$^{-1}$. 


Only with additional epochs will we be able to test if the above solution predicts the correct $v(r)$ and $a(r)$ for the outflow. Based on the above derived outflow parameters, we integrate equation (\ref{eq:vr2}) and predict the accelerations of S4 for the next 2, 5, and 10 years in the observed frame (see table \ref{tb:ACCpredict}). Under the disk-wind model, the outflow's acceleration decreases by $\sim$ 40\% and the velocity reaches $\simeq$ 22,500 km s$^{-1}$ in 10 years. These results are similar to the BAL outflow acceleration reported in \cite{Grier16}. They observed \civ\ BAL accelerations between three epochs for quasar SDSS J0124--0033. The observed average acceleration dropped from 0.90 cm s$^{-2}$ between epochs 1 and 2 to 0.37 cm s$^{-2}$ between epochs 2 and 3. However, they found that the disk-wind model is insufficient to explain their observations since the parameters derived from their epochs 1 and 2 overpredict the velocity shift by about a factor of five when applied to epoch 3. 

Similarly, additional HST/COS observations of quasar SDSS J1042+1646 in the next decade will be able to test the prediction of the radiatively-driven disk-wind model (see table \ref{tb:ACCpredict}). In addition, for outflow S4, we have distance constraints (see section \ref{sec:ne}). Therefore, contrasting the model predictions with the $r$, $v$, and $a$ extracted from future observations will be particularly instructive for testing and understanding the acceleration mechanisms of quasar outflows.

\section{Summary}

In this paper, we identified and analyzed the BAL acceleration for outflow S4 in quasar SDSS J1042+1646. The main results are summarized as follows:\\

1. We observed significant velocity shift signatures in multiple ionic absorption troughs for outflow system 4. The \neviii\ absorption troughs show similar velocity structures over the six-year interval (see section \ref{section:acc}), while the trough centroids shifted by -1550 km s$^{-1}$ over 3.2 yr in the quasar rest frame. Moreover, for both the 2011 and 2017 epochs, we obtained the photoionization models using the Synthetic Spectral Simulation method and the photoionization solutions are similar for the two epochs (see section \ref{sec:SSS}). These two points support the claim that we observe the same outflow but it is shifted by --1550 km s$^{-1}$.

2. We attribute the velocity shift to acceleration since we are able to exclude time-dependent photoionization changes and motion of material into and out of the LOS as alternate explanations (see section \ref{sec:BALacc}). This leads to an average acceleration of 480$^{+50}_{-50}$ km s$^{-1}$ yr$^{-1}$ or 1.52$^{+0.16}_{-0.16}$ cm s$^{-2}$. 

3.  We compared our results with previous studies of BAL accelerations and concluded that the S4 outflow has the largest velocity shift and acceleration observed in BAL outflows to date (see section \ref{sec:Other}). This is also the first time where quasar outflow acceleration is observed from more than one ion and in distinct troughs from a doublet transition (\neviii, see section \ref{section:acc}).

4. The outflow velocity and \Nh\ are similar to the high-velocity X-ray outflow reported in PG 1211+143, which suggest that we probe similar outflows in both cases (see section \ref{sec:UFO}).

5. Using the observed velocities and associated acceleration, the disk-wind model of \cite{Murray97} yields $R$ = 0.23 pc for outflow S4. We also have distance constraints derived from the \ov*\ multiplet (0.05 pc $<$ $R$ $<$ 54.3 pc, see section \ref{sec:ne}). The disk-wind model makes predictions for future values of $v$, $a$, and $r$ (see table \ref{tb:ACCpredict}), which can be uniquely tested with future HST/COS observations (see section \ref{sec:DiskWind}).

\acknowledgments
X.X., N.A., and T.M acknowledge support from NSF grant AST 1413319, as well
as NASA STScI grants GO 14777, 14242, 14054, and 14176, and NASA ADAP 48020.

Based on observations made with the NASA/ESA \textit{Hubble Space Telescope}, and obtained from the data archive at the Space Telescope Science Institute. STScI is operated by the Association of Universities for Research in Astronomy, Inc. under NASA contract NAS5-26555.


\bibliography{apj-jour,dsr-refs}

\end{document}